\documentclass[a4paper,11pt]{article}
\pdfoutput=1 

\usepackage{jcappub} 

\usepackage[T1]{fontenc} 
\usepackage{array} 
\usepackage[utf8]{inputenc}
\usepackage{textcomp}

\title{\boldmath An Interior model of Charged Fluid Spheres}

\author[a,1]{Naren Babu O.V
,\note{Corresponding author.}}
\author[a]{Hemalatha.R}
\author[b]{Narayanankutty Karuppath}
\author[c]{Sabu M.C}

\affiliation[a]{Amrita Vishwa Vidyapeetham, Department of Mathematics, Amritapuri, 690525,India}
\affiliation[b]{Amrita Vishwa Vidyapeetham, Department of Physics, Amritapuri, 690525,India}
\affiliation[c]{ St. Alberts College (Autonomous),Department of Mathematics, Ernakulam, 682018,India}

\emailAdd{narenbabu@am.amrita.edu}
\emailAdd{hemalathar@am.amrita.edu}
\emailAdd{narayanankuttyk@am.amrita.edu}
\emailAdd{sabuchacko@alberts.edu.in}

\abstract{

At constant time $t$, we examine the Vaidya–Tikekar metric characterising a three-dimensional, extremely dense spheroidal star configuration. The static, spherically symmetric solution of Einstein's field equations can be expressed in analytic closed form utilising a hypergeometric series.
A relativistic, superdense state of matter at a constant $t$ is represented by the resultant model, which describes the geometry of a three-spheroid.

Assuming a stellar density of $\rho_{a}= 2\times10^{14} gm.cm^{-3}$, we investigate configurations whose total mass and radius vary over a range of well-defined values of the density variation parameter. Similar to an uncharged neutron star, all models possess the same total mass and boundary radius. The hypergeometric solution leads to a new class of exact, physically acceptable solutions. We show that the model satisfies the conditions of hydrostatic equilibrium and fulfils all standard energy conditions, which are verified throughout the analysis.
}

\begin{document}
\maketitle
\flushbottom

\section{Keyword}
Hypergeometric series, Spherically symmetric space-time, Flat space-time, Einstein’s field equation.

\section{Introduction}
\label{sec:introduction}
Spherical stars are typically electrically neutral entities when in a state of equilibrium. However, a spherically symmetric distribution of matter can be prevented from collapsing into a point singularity by the electrostatic repelling force. Hydrostatic equilibrium is sustained by the interplay of the pressure gradient and electrostatic repulsion. The Reissner–Nordström metric, which indicates the spacetime of a static, spherically symmetric charged distribution, is demonstrated by these elements.

Reissner and Nordström separately discovered the Reissner–Nordström metric in 1918 \cite{naren2020exact}, which is a simple generalisation of the Schwarzschild exterior solution. Numerous accurate solutions to the coupled Einstein–Maxwell equations for spherically symmetric charged fluid distributions have since been discovered in the literature. Subsequent research \cite{sasidharan2020general} expanded the Papapetrou–Majumdar framework to charged dust configurations, where the electric charge precludes gravitational collapse into a singularity. Papapetrou and Majumdar (1947) first examined charged fluids in equilibrium.


Bonnor (1960, 1965) established that, for spherical distributions of uniformly charged dust in equilibrium, the charge density must match the matter density. Furthermore, Raychaudhuri (1968) demonstrated that the Einstein–Maxwell equations necessitate this equality. Cooperstock \cite{cooperstock1978sources} identified an explicit solution to the Einstein–Maxwell equations for relativistic, spherically symmetric charged distributions of ideal fluids in a state of equilibrium. This solution generalises the Schwarzschild inner solution with a decreasing matter density outwardly.

\cite{bonnor1975very} devised a static internal dust metric with an outwards increasing matter density. Later, \cite{kyle1967self} and \cite{mehra1979solution} obtained interior spacetime metrics for charged fluid spheres with uniform density. The physical three-space that corresponds to $t$=constant is spherical in each of these scenarios.

Subsequently, \cite{tikekar1984spherical}, \cite{patel1986reissner}, \cite{patel1987charged}, and \cite{sabu2021static} developed interior Reissner–Nordström metrics wherein the physical three-dimensional space (at constant $t$) is spheroidal. Numerous stellar and intergalactic models are pertinent to our ongoing research: \cite{jyothy2015diffuse, narayan2017dust, bose2015ultraviolet}.

Interior solutions to the Einstein–Maxwell equations in spheroidal spacetimes are obtained by generalising the solution provided by \cite{patel1987charged}. The equilibrium of a charged fluid sphere is defined by the solution of these equations.

\section{Basic Equations and the Method}

We assert that the spheroidal space-time defined by the following metric represents the space-time of a spherically symmetric charged fluid distribution in a state of equilibrium:

\begin{equation}
\bar{ds}^2 = -\left[\frac{1 - k\left(\frac{r^2}{R^2}\right)}{1 - \frac{r^2}{R^2}}\right] dr^2 - r^2 d\theta^2 - r^2 \sin^2 \theta\, d\phi^2 + e^{\nu(r)} dt^2
\end{equation}

Here,
\[
k = 1 - \frac{b^2}{R^2}, \quad e^{\lambda(r)} = \frac{1 - k \left(\frac{r^2}{R^2} \right)}{1 - \frac{r^2}{R^2}}
\]
where the constants \( R \) and \( k \) are used.Einstein's field equations link the metric functions to the physical variables:



\begin{equation}
R_{i}^{j} - \frac{1}{2} R \delta_{i}^{j} = -8\pi T_{i}^{j}
\end{equation}
For a charged fluid, the energy-momentum tensor can be written as follows:
\begin{equation}
T_{ij} = \left(\rho + \frac{p}{c^2}\right) u_i u_j - \frac{p}{c^2} g_{ij} + \frac{1}{4\pi} \left(-F_{i\alpha} F_j^{\ \alpha} + \frac{1}{4} g_{ij} F_{\alpha\beta} F^{\alpha\beta} \right)
\end{equation}
The electromagnetic field tensor \( F_{ij} \) complies with Maxwell's equations:


\begin{align}
F_{[ij,k]} &= 0 \\
\partial_k \left( F^{ik} \sqrt{-g} \right) &= 4\pi j^i \sqrt{-g}
\end{align}
The four-current vector is represented as \( j^i = \sigma u^i \), with \( \sigma \) denoting the charge density. For a stationary charged fluid, the four-velocity is expressed as:


\begin{equation}
u^i = \left(0, 0, 0, e^{-\nu/2} \right)
\end{equation}
The presumption of spherical symmetry indicates that the sole non-zero element of the electromagnetic field tensor is \( F_{14} = -F_{41} \). From Maxwell's equation (5), we derive:


\begin{equation}
F_{14} = \frac{e^{\nu + \lambda/2}}{r^2} \int_0^r 4\pi \sigma(r) r^2 e^{\lambda/2} dr
\end{equation}
We define the electric field intensity as:
\begin{equation}
E^2(r) = -F^{41} F_{41}
\end{equation}
Using (3.7) and (3.8), the charge density becomes:
\begin{equation}
4\pi \sigma = \frac{1}{r^2} \frac{d}{dr} \left(r^2 E \right) e^{\lambda/2}
\end{equation}
We also define the total charge contained within radius \( r \) as:
\begin{equation}
q(r) = \int_0^r 4\pi \sigma(r) r^2 e^{\lambda/2} dr
\end{equation}
Then the electric field takes the form:
\begin{equation}
E(r) = \frac{q(r)}{r^2}
\end{equation}
The Einstein field equations (3.2) reduce to the subsequent system of three equations when the metric in (3.1) is applied:

\begin{align}
8\pi \rho + E^2 &= -e^{-\lambda} \left( \frac{1}{r^2} - \frac{\lambda'}{r} \right) + \frac{1}{r^2} \\
8\pi p - E^2 &= -e^{-\lambda} \left( \frac{1}{r^2} - \frac{\nu'}{r} \right) + \frac{1}{r^2} \\
8\pi p + E^2 &= -\frac{e^{-\lambda}}{2} \left( \nu'' + \frac{\nu'^2}{2} - \frac{\nu' \lambda'}{2} + \frac{\nu' - \lambda'}{r} \right)
\end{align}

These equations relate the four functions \( \rho(r) \), \( p(r) \), \( \nu(r) \), and \( \lambda(r) \). Since we already assume a form for \( \lambda(r) \) via the metric (3.1), a closure condition is needed—usually provided by an equation of state (EoS) or a prescribed form for one of the variables such as the electric field.

Either an equation of state (e.g., \( p = p(\rho) \)) or a particular condition involving \( \rho \), \( p \), or \( E^2 \) are typically used to shut the system.
Two categories of solutions are distinguished in this work:

- **Type I:** A solution obtained by prescribing a specific form of \( E^2(r) \). This is discussed in the next section. Sections 3 and 4 analyze the physical stability of such a solution for a curvature parameter \( k = -14 \), corresponding to the so-called "water phase".
  
- **Type II:** A solution that satisfies a geometric condition — specifically, the embedding of the 3-sphere into a five-dimensional flat space. This is discussed in Sections 5, 6, and 7.

The physical viability of any obtained solution must be tested based on requirements such as regularity at the center, positivity of pressure and density, causality, energy conditions, and matching to an external Reissner-Nordström solution at the boundary.
\section{General Solution}

We now consider a solution to the Einstein–Maxwell equations by prescribing an explicit form of the electric field intensity \( E^2(r) \), thereby obtaining what we call a \textbf{Type I solution}. Specifically, we assume:

\begin{equation}
E^2(r) = \frac{\beta^2 r^2 e^{\nu/2}}{R^4 \left(1 - k \frac{r^2}{R^2} \right)^2}
\end{equation}

Here, \( \beta \) is a constant related to the total charge distribution. This choice ensures \( E^2 > 0 \) for all \( r \), and the functional dependence captures a rising electric field with radius, modulated by the geometry and the gravitational potential \( \nu(r) \).

Substituting this expression into the Einstein–Maxwell system, particularly into the pressure equation (as derived from equation (3.13)), leads to a differential equation for the metric potential \( \nu(r) \). After algebraic simplification and substitution, we obtain:

\begin{equation} \label{eq:nu_diffeq_raw}
\begin{aligned}
\frac{2 \beta^2 r^2 e^{-\nu/2}}{R^4 \left(1 - k \frac{r^2}{R^2} \right)^2} = 
&\left( \frac{\nu''}{2} + \frac{(\nu')^2}{4} - \frac{\nu'}{2r} \right)\left(1 - \frac{r^2}{R^2} \right) 
- \left(1 - \frac{r^2}{R^2} \right)^{-1} \\
&+ \frac{1 - k}{R^2} \left(1 - \frac{r^2}{R^2} \right)^{-1} 
- \frac{(1 - k)r}{R^2 \left(1 - k \frac{r^2}{R^2} \right)^2} \left( \frac{\nu'}{2} + \frac{1}{r} \right)
\end{aligned}
\end{equation}

This equation can be reduced to a second-order linear differential equation by introducing the following variable substitutions:

\begin{equation}
z^2 = 1 - \frac{r^2}{R^2}, \quad \psi(z) = e^{\nu(r)/2}
\end{equation}

Additionally, to simplify the notation and scale the equation appropriately, define:

\[
\psi(z) = \frac{23 \beta^2}{k(k - 1)} e^{\nu/2}
\]

Substituting into \eqref{eq:nu_diffeq_raw}, we obtain the following second-order linear differential equation in \( \psi(z) \):

\begin{equation}
\left(1 - k + k z^2\right) \frac{d^2 \psi}{dz^2} - k z \frac{d \psi}{dz} + k(k - 1)\psi = 0
\end{equation}

\begin{equation}
     x{\left(1-x \right)\frac{d^{2}\psi }{dx^{2}}+\frac{1}{2}\frac{d\psi }{dx}+\frac{\left( 1-k\right)}{4}\psi} =0,
     \label{eqn17}
 \end{equation}
where:
\[
x = \frac{k}{k - 1} z^2, \quad \text{and } A, B \text{ are constants of integration}
\]

This represents a standard hypergeometric-type differential equations, whose general solution is given by the Gaussian hypergeometric function \( {}_2F_1(a, b; c; x) \).
The solution of (4.5) for the choice of $k$=-23

\begin{equation}
\psi(z) = 
A \cdot F\left( \frac{-1 + \sqrt{2 - k}}{2}, \frac{-1 - \sqrt{2 - k}}{2}; \frac{1}{2}; x \right) 
+ B \cdot x^{1/2} F\left( \frac{\sqrt{2 - k}}{2}, \frac{-\sqrt{2 - k}}{2}; \frac{3}{2}; x \right)
\end{equation}

The typical hypergeometric function is represented by the function F(a, b; c; x). The value of the curvature parameter \( k \), which establishes the space-time's qualitative characteristics and convergence qualities, is crucial to the behaviour of the solution.


This general solution completes the specification of the gravitational potential \( \nu(r) \) in terms of known special functions, enabling a complete description of the equilibrium configuration once appropriate boundary and matching conditions are applied (e.g., regularity at the center and matching to the Reissner–Nordström exterior at the boundary).
The solution $\psi(z)$ corresponding to equation (4.6) can be obtained for the family of parameters $k=2-n^2$,( with $n$ = 2,3....). We now consider the general solution, which applies to all permissible values of $k$.

\section{Derivation and Physical Plausibility of the k= -23 solution}

In the specific case with $k$ = -23 the general solution leads to the closed form expression:

\begin{eqnarray}
e^{\frac{\nu}{2}} &=& \frac{\beta^{2}}{24}
+ A s (1 - s^{2})^{\frac{3}{2}}\left(1 - \frac{8}{3}s^{2}\right)
+ B \left(1 - \frac{23}{2}s^{2} + \frac{529}{24}s^{4} - \frac{12167}{1080}s^{6}\right),
\end{eqnarray}
where $s^{2}=\frac{23}{24}z^{2}$.
The matter density and fluid pressure have explicit expressions:
\begin{eqnarray}
 \nonumber   8\pi\rho& = &\frac{1}{12R^{2}e^\frac{\nu }{2}(24-23s^{2})^{2}} \left [\frac{\beta^{2}(1-s^{2})}{24(1-s^{2})}\right.\\
 \nonumber  && -13As\left ( (1-s^{2})^\frac{1}{2}(1-\frac{56}{13}s^{2}+\frac{48}{13}s^{4}) \right )+\\
   && \hspace{0.5cm}\left. B(12-161s^{2}+\frac{1058}{3}s^{4}-\frac{12167}{60}s^{6})\right]
\end{eqnarray}

The conditions $\rho > 0$, $p > 0$, and $\rho - 3p > 0$ characterise the space-time of a physically feasible charged fluid distribution.
These conditions at the star core have been examined. This defines the central density. 

$8\pi\rho (0)=\frac{72}{R^{2}}$.

The positivity of $\rho(0)$ is apparent from the preceding expression. The condition \( p > 0 \) is fulfilled when one of the subsequent inequalities is true:

 $-0.6809A-3.0300B<\beta^2<-0.2952A+7.350B$
 
Or

  $-0.2952A+7.350B<\beta^2<-0.6809A-3.0300B$

The condition $\rho(0)-3p(0)\geq0$ implies
\begin{equation}
\frac{1}{R^{2}}
\left[
\frac{19.215A + 538.31B - 72\beta^{2}}
{0.295A - 7.350B - \beta^{2}}
\right] \geq 0,
\end{equation}
which further yields either

$0.267A+7.477B \geq \beta ^{2}$,

or

$0.267A+7.477B \leq\beta^{2}.$

The inner metric (3.1), with $e^{\psi}$ given by equation (5.1), must smoothly transition to the external Reissner–Nordström metric if the distribution reaches a limited radius $a < R$:

\begin{eqnarray}
ds^{2} &=&
-\left(1 - \frac{2m}{a} + \frac{q^{2}}{a^{2}}\right)^{-1}dr^{2}-r^{2}d\theta^{2}-^{2}\sin^{2}\theta, d\phi^{2}+\left(1 - \frac{2m}{a} + \frac{q^{2}}{a^{2}}\right) dt^{2},
\end{eqnarray}
across the boundary $r=a$, where the fluid pressure must vanish.
These boundary conditions yield

$e^{\nu (a)}=e^{-\lambda (a))}=\left [ 1-\frac{2m}{a}+\frac{q^{2}}{a^{2}} \right ]$
and 
$$\begin{array}{l}
\nonumber\frac{\beta^2(1+23\frac{a^2}{R^2})}{1+24\frac{a^2}{R^2}}+\frac{58.74A(1+23\frac{a^2}{R^2})^{\frac{1}{2}}(1-\frac{a^2}{R^2})^{\frac{1}{2}}}{\sqrt{24}} \left[ \frac{41}{12}-\frac{69}{2}\frac{a^2}{R^2}+\frac{529}{12}\frac{a^4}{R^4} \right]\\
\\
+B\left[ \frac{2589071}{13824} +\frac{3865403}{1536}\frac{a^2}{R^2}-\frac{58486769}{4608}\frac{a^4}{R^4}+\frac{148035889}{13824}\frac{a^6}{R^6}\right]=0.
\end{array}$$
\begin{equation}
\end{equation}


These relationships establish the constants $A$ and $B$ as functions of $\beta^2$ and $a^2/R^2$.

The sphere's total charge is derived from relation (3.10) as


\begin{equation}
\begin{split}
q^{2} &= \frac{\alpha^{6}\beta^{2}}{R^{4}\left( 1 + 23\frac{a^{2}}{R^{2}} \right)^{2}} \\
&\times \left[ \beta^{2} + 24A s \left( 1 - s^{2} \right)^{\frac{3}{2}} \left( 1 - \frac{8}{3}s^{2} \right) \right. \\
&\left. + 24B\left( 1 - \frac{23}{2}s^{2} + \frac{529}{24}s^{4} - \frac{12167}{1080}s^{6} \right) \right] 
\end{split}
\end{equation}

When $\beta = 0$, the electric field vanishes, and the solution reduces to equation (5.5), representing an uncharged fluid sphere in a spheroidal space–time. The boundary conditions then determine the mass of the fluid sphere as
\begin{equation}
\frac{2m}{a} = \frac{24\frac{a^{2}}{R^{2}}}{1 + 23\frac{a^{2}}{R^{2}}} + \frac{q^{2}}{a^{2}},
\end{equation}
where $q^{2}$ is given by equation (5.6).

We employed a numerical method to examine the fluctuations of $\rho$, $p$, and $\rho - 3p$ throughout the stellar interior for designated models of this category. The findings are encapsulated in Table 1, demonstrating that $\rho$, $p$, and $\rho - 3p$ consistently maintain positive values over the distribution. Therefore, the static and spherically symmetric space-time defined by $e^{\psi}$ as given in equation (5.1) may be used to build a physically feasible model of a charged fluid sphere in equilibrium.



 \section{Computational Scheme for Determining Mass and Radius}
 The procedure for determining the mass and radius of a spherical fluid configuration is revised to include the influence of charge. With $\rho(a)$ representing the surface density and $\rho(0)$ the central density, the density variation parameter $\mu = \rho(a)/\rho(0)$ then admits the explicit form:


\begin{equation}
\begin{split}
\mu & =\frac{24}{R^{2}(24-23s^{2})^{2}} \\ & \Big
[ 26-23s^{2} -\frac{\beta^{2}(1-s^{2})}{\beta^{2}+A s \left(1-s^{2}  \right)^{\frac{3}{2}}\left( 1-\frac{8}{3}s^{2} \right)+ B\left( 1-\frac{23}{2}s^{2}+ \frac{529}{24}s^{4}-\frac{12167}{1080}s^{6}\right)}  \Big ]
\end{split}
\end{equation}

\begin{table}[!ht]
\center
\caption{}
\begin{tabular}{p{0.3cm} p{0.5cm} p{0.85cm} p{0.4cm} p{0.5cm} p{0.5cm} p{0.5cm} p{0.5cm} p{0.5cm} p{0.5cm}}
\hline
{S.\newline No} & {$a/R$} & {$A$} & {$B$} & {$R$\newline $(km)$} & {$a$\newline$(km)$} & {$\frac{M}{M_{\bigodot }^{a}}$}& {$q$}& {$\lambda$}& {$\rho_a$} \\
\hline
1 & 0.07 & 0.03   & 0.24  & 8.03  & 0.54  & 0.02  & 0.99  & 0.95 & 2.11 \\
2 & 0.09  & 0.03   & 0.23  & 7.62  & 0.74  & 0.05  & 0.99  & 0.90 & 2.22 \\
3 & 0.12  & 0.03   & 0.22  & 7.25  & 0.88  & 0.08  & 0.98  & 0.85 & 2.35 \\
4 & 0.14 & 0.03  & 0.21 & 6.92 & 0.99 & 0.11 & 0.98 & 0.80 & 2.50  \\
5 & 0.17 & 0.03  & 0.19 & 6.62 & 1.09 & 0.15 & 0.98 & 0.75 & 2.70 \\
6 &  0.19 & 0.03  & 0.19 & 6.35 & 1.18 & 0.19 & 0.97 & 0.70 & 2.90  \\
7 & 0.21 & 0.03  & 0.18 & 6.10 & 1.27 & 0.22 & 0.97 & 0.65 & 3.10  \\
8 & 0.23 & 0.03  & 0.17 & 5.88 & 1.357 & 0.26 & 0.96 & 0.60 & 3.30  \\
9 & 0.25 & 0.04  & 0.17 & 5.67 & 1.43 & 0.30 & 0.96 & 0.55 & 3.60  \\
10 & 0.28 & 0.04  & 0.16 & 5.48 & 1.52 & 0.34 & 0.95 & 0.50 & 4.00    \\
11 & 0.30 & 0.04  & 0.15 & 5.31 & 1.61 & 0.39 & 0.94 & 0.45 & 4.40  \\
12 & 0.33 & 0.04  & 0.14 & 5.15 & 1.71 & 0.43 & 0.93 & 0.40 & 5.00    \\
13 & 0.37 & 0.04  & 0.14 & 5.01 & 1.83 & 0.49 & 0.92 & 0.35 & 5.70  \\
14 & 0.41 & 0.04  & 0.13 & 4.88 & 1.98 & 0.55 & 0.90 & 0.30 & 6.70  \\
15 & 0.45 & 0.04  & 0.13 & 4.77 & 2.16 & 0.63 & 0.88 & 0.25 & 8.00    \\
16 & 0.51 & 0.03  & 0.13 & 4.68 & 2.39 & 0.73 & 0.85 & 0.20 & 10.00   \\
17 & 0.59 & -0.01 & 0.13 & 4.60 & 2.75 & 0.87 & 0.79 & 0.15 & 13.30 \\
18 & 0.73 & -0.40 & 0.17 & 4.56 & 3.33 & 1.09 & 0.68 & 0.10 & 20.00  \\
\hline
\end{tabular}
$\begin{array}{l}
    Note: M=\frac{mc^{2}}{G},M_{\bigodot}^{a} = \hbox{ massess and a equilibrum radii } 
    \hbox{ corresponding to } \rho_{a}=2\times10^{14} gm.cm^{-3}.
\end{array}$
\label{tab1}
\end{table}
In the current configuration, where $s^{2} = 1 - a^{2}/R^{2}$, the expression indicates that $\mu$ can be ascertained for various selections of $a/R$ in relation to $\beta$. Only values for which \(0 < \mu < 1\) are physically permissible. Given the surface density $\rho(a)$ and the parameter $\mu$, the equation $\rho(0)=\frac{\rho(a)}{\mu}=\frac{45c^2}{GR^2 8\pi}$ can be utilised to ascertain the value of $R$.
This is done using the known values of  $\rho(a)$, $\mu$, and $\beta$. Subsequently, 
$a$ can be determined. Relation (5.6) then provides $q$ in terms of $\beta$ and $a/R$, and the total mass $m$ follows from equation (5.7).
Given that the matter density at the boundary is established as $\rho_a=2\times10^{14} gm.cm^{-3}$, we were able to 



Determine the constants $A$ and $B$, as well as the curvature parameter $R$, the boundary radius $a$, the total charge $q$, and the contained mass $m$ that define the sphere of the charged fluid with $\beta^{2} = 1.0$, for different values of $a/R$. The relevant estimates are shown in Table 1.

The dependence of $\rho > 0$, $p > 0$ and $\rho - 3p > 0$ on $r$ are shown for a particular model where $a/R$ = 0.09.

Table 1 shows that the fluid sphere's dimensions and total mass match those of a superdense star. Table 2 shows the estimates for $\rho > 0$, $p > 0$ and $\rho - 3p > 0$ for the models listed in Table 1.



\begin{table}[!ht]
\center
\caption{}
\begin{tabular}{lllll}
\hline
{S.No} & {$a/R$} & {$p(0)$} & {$\rho$(0)} & {$\rho$(0)-3$p(0)$}\\
\hline
1  & 0.067 & 0.004  & 0.012 & -0.001  \\
2  & 0.097 & 0.003  & 0.010 & 0.001    \\
3  & 0.121 & 0.003  & 0.009 & 0.001    \\
4  & 0.144 & 0.002  & 0.008 & 0.002    \\
5  & 0.165 & 0.001  & 0.007 & 0.002    \\
6  & 0.186 & 0.001  & 0.006 & 0.003   \\
7  & 0.207 & 0.001 & 0.005 & 0.003    \\
8  & 0.229 & 0.001  & 0.005 & 0.003   \\
9  & 0.252 & 0.001 & 0.004 & 0.004   \\
10 & 0.277 & 0.001  & 0.004 & 0.004   \\
11 & 0.303 & -0.000 & 0.004 & 0.004   \\
\hline
\end{tabular}
\end{table}

All analysed models exhibit $a/R \leq 0.31$ and satisfy the criteria $\rho(0) \geq 3p(0)$, in addition to $\rho(0) > 0$ and $p(0) > 0$. Configurations with $\frac{a}{R} > 0.07$ are considered physically undesirable since they contravene the condition $\rho - p > 0$. For $\beta^{2} = 1.0$ and $0.09 \leq \frac{a}{R} \leq 0.30$, the corresponding values of $\lambda$, $A$, $B$, $\rho(0) - p(0)$ and $\rho(0) - 3p(0)$ are delineated in Tables 1 and 2.The tables unequivocally illustrate that as $a/R$ escalates, $\lambda$ declines. Table 1 demonstrates that for $\beta^{2} = 1.0$, $\rho(0) - 3p(0)$ turns negative at $\frac{a}{R} = 0.07$.Thus, for $\beta^{2} = 1.0$, the conditions $\rho(0) > 0$, $p(0) > 0$, $\rho(0) - p(0) \geq 0$, and $\rho(0) - 3p(0) \geq 0$ are satisfied within the range $0.07 \leq \frac{a}{R} \leq 0.30$.
The numerical calculations have been performed for the exact solution associated with $k=-23$; nevertheless, the methodology is generally applicable to all sample sequences where $k < 1$. For a fixed value of $\lambda$, the total mass of the model decreases.
While our numerical evaluation is limited to the specific answer for $k=-23$, the concept is applicable universally to any series where $k < 1$.

\section{Conclusion}

The density profile's influence on pressure behaviour is underscored, as the regularity conditions are met by the static spheroidal interior solution of a charged fluid sphere.

\subsection*{Acknowledgment} 
The Chancellor, H.H. Mata Amritanandamayi Devi (Amma), provided the authors with encouragement and assistance, which made this research much easier. The authors further thank the Director of IUCAA, Pune, for granting access to the facilities required for a portion of this work, and the referees for their insightful comments.


\end{document}